\documentclass[preprint]{aastex62}
\usepackage{amsmath,amssymb,CJK,soul}

\received{--}
\revised{--}
\accepted{--}
\submitjournal{ApJL}

\shorttitle{Recent Disruption of (3200) Phaethon}
\shortauthors{Ye, Wiegert, \& Hui}

\begin{document}
\begin{CJK*}{UTF8}{gbsn}

\title{In Search of Recent Disruption of (3200) Phaethon: Model Implication and Hubble Space Telescope Search}

\correspondingauthor{Quanzhi Ye}
\email{qye@caltech.edu}

\author[0000-0002-4838-7676]{Quanzhi Ye (叶泉志)}
\affiliation{Division of Physics, Mathematics and Astronomy, California Institute of Technology, Pasadena, CA 91125, U.S.A.}
\affiliation{Infrared Processing and Analysis Center, California Institute of Technology, Pasadena, CA 91125, U.S.A.}

\author{Paul A. Wiegert}
\affiliation{Department of Physics and Astronomy, The University of Western Ontario, London, Ontario N6A 3K7, Canada}
\affiliation{Centre for Planetary Science and Exploration, The University of Western Ontario, London, Ontario N6A 5B8, Canada}

\author{Man-To Hui (许文韬)}
\affiliation{Department of Earth, Planetary and Space Sciences, UCLA, Los Angeles, CA 90095-1567, USA}



\begin{abstract}

Near-Earth asteroid (3200) Phaethon is notable for its association to a strong annual meteor shower, the Geminids, indicative of one or more episodes of mass ejection in the past. The mechanism of Phaethon's past activity is not yet understood. Here we present a Hubble Space Telescope (HST) search of meter-sized fragments in the vicinity of Phaethon, carried out during Phaethon's historic approach to the Earth in mid-December of 2017. Numerical simulation conducted to guide HST's pointing also show that the dynamical evolution of Phaethon-originated particles is quick, as ejected materials take no longer than $\sim250$~yr to spread to the entire orbit of Phaethon. Our search was completed down to 4-meter-class limit (assuming Phaethon-like albedo) and was expected to detect 0.035\% particles ejected by Phaethon in the last several decades. The negative result of our search capped the total mass loss of Phaethon over the past few dozen orbits to be $10^{12}$~kg at $3\sigma$ level, taking the best estimates of size power-law from meteor observations and spacecraft data. Our result also implies a millimeter-sized dust flux of $<10^{-12}~\mathrm{m^{-2}~s^{-1}}$ within 0.1~au of Phaethon, suggesting that any Phaethon-bound mission is unlikely to encounter dense dust clouds.

\end{abstract}

\keywords{minor planets, asteroids: individual [(3200) Phaethon] --- meteorites, meteors, meteoroids}


\section{Introduction} \label{sec:intro}

Near-Earth asteroid (3200) Phaethon is dynamically associated with the strong Geminid meteor shower \citep{Whipple1983,Williams1993}, as well as several other asteroids \citep{Ohtsuka2006,Kasuga2009}, collectively known as the Phaethon-Geminid Complex (PGC). It has long been known that strong meteor showers are typically associated with unambiguous comets which activities are driven by sublimation of cometary water ice. However, numerous observations of Phaethon taken in the past several decades have so far rejected Phaethon as a typical comet \citep[e.g.][]{Cochran1984,Chamberlin1996,Hsieh2005,Licandro2007,Wiegert2008,Jewitt2010,Jewitt2013,Li2013,Hui2017}. The formation mechanism of PGC remains an intriguing question.

One peculiar aspect of Phaethon is its orbit: Phaethon has an orbital period of 1.4~yr and a perihelion distance of $q=0.14$~au. This leads to its frequent exposure to extreme solar heating. Recent work by \citep{Granvik2016} suggested that asteroids with small perihelion distances such as Phaethon are prone to catastrophic disruptions due to the extensive thermal stress they experience. Curiously, the behavior of the PGC system -- several asteroids and a dense meteoroid stream being dynamically related to each other -- is in line with a disintegrative origin. To examine this hypothesis, it is important to know how materials got ejected and how they evolve.

Phaethon has been selected as the target for the DESTINY$^+$ mission (abbreviation of ``Demonstration and  Experiment of Space Technology for INterplanetary voYage, Phaethon fLyby and dUst Science Phaethon fLyby with reUSable probe''), currently being considered by the Japan Aerospace Exploration Agency (JAXA). One of the mission goals is to understand the dust environment in Phaethon's vicinity \citep[e.g.][]{Krueger2017, Arai2018}. Since dust grains in Phaethon's vicinity will predominately be young ejecta from Phaethon, it is useful to understand Phaethon's recent activity. However, this is a challenging task as Phaethon's current activity is confined to the perihelion, which only become accessible after the launch and operation of the Solar and Terrestrial Relations Observatory (STEREO) in 2006. Despite on an Earth-approaching orbit, Phaethon does not approach the Earth often, with the last close ($<0.1$~au) approach in 1974, making it difficult to study anything in its vicinity.

On 2017 December 16, Phaethon passed only 0.07~au from the Earth, the closest since 1974 and also until 2093. This close approach provides an excellent and rare opportunity to study young ejecta in Phaethon's vicinity. Here we report our modeling and observational investigation of such ejecta, as well as its implication on the recent behavior of Phaethon.

\section{Dynamical Evolution of Phaethon's Recent Ejecta} \label{sec:sim}

The dynamics of Phaethon's ejecta is primarily controlled by gravitational attraction from the Sun and major planets, and, for the case of smaller particles, radiation pressure exerted by sunlight. The magnitude of the exerted radiation pressure is typically parameterized by $\beta$, which is inversely proportional to the product of particle density $\rho$ and size $r$, $\beta \propto (\rho r)^{-1}$. Meteoroids visible to naked eye, optical and radar instruments typically have $\beta\sim0.001$, and $\beta\rightarrow 0$ for particles at meter sizes or larger. All these effects, gravitational and radiational, will cause Phaethon's ejecta to slowly drift away from Phaethon. The timescale of such dispersion can be understood through dynamical modeling, since the dynamics of gravitational perturbations and radiation effects are well known. The major unconstrained process in this model is the ejection velocity itself, which we take as a free parameter.

We use a MERCURY6-based \citep{Chambers2012} package developed in our earlier works \citep[e.g.][]{Ye2015,Ye2016a} for our investigation. Since the ejection mechanism is not clearly known, we carry out three separate simulations, one with the conventional Whipple cometary ejection model \citep{Whipple1950} for $\beta=0.001$ particles, the other two with gravitational escape ejection speed for both $\beta=0$ and $\beta=0.001$ particles. (Whipple's cometary model cannot eject $\beta=0$ particle: the ejection speed is below gravitational escape speed.) For reader's reference, the gravitational escape speed is $\sim3~\mathrm{m/s}$ which can be derived from Phaethon's mass assuming a spherical body, with diameter of $5.1$~km \citep{Hanus2016} and a density of $2900~\mathrm{kg/m^3}$ \citep{Babadzhanov2009}; the ejection speed of $\beta=0.001$ particle under cometary mechanism is $\sim16~\mathrm{m/s}$. Particles are isotropically released from Phaethon upon the perihelion of 2002 May 2 (chosen arbitrarily) and are integrated using a Bulirsch-Stoer integrator, considering the gravitational perturbations from the eight major planets (with the Earth-Moon system represented by a single point mass at the barycenter of the two bodies), relativistic effects, and for $\beta=0.001$ particles, radiation pressure and Poynting-Robertson drag. The particle representing Phaethon and all dust particles are considered to be massless and do not interact with each other.

\begin{figure*}
\includegraphics[width=\textwidth]{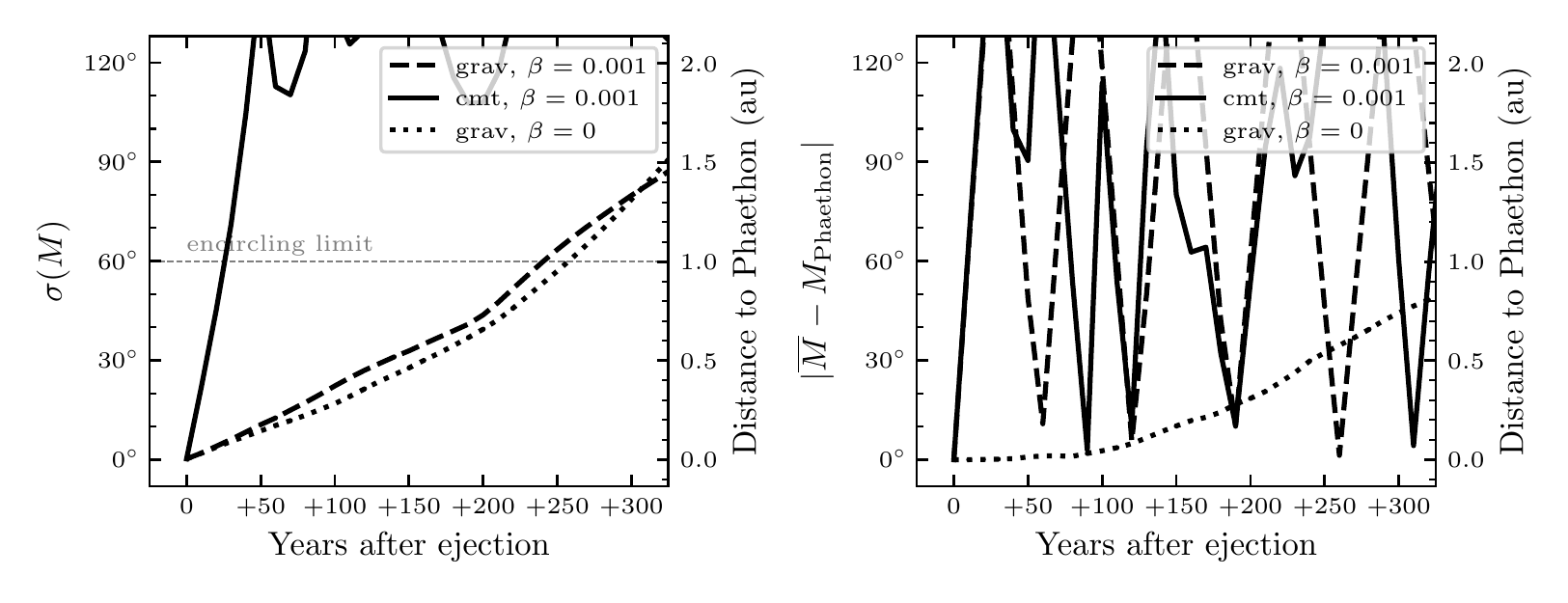}
\caption{Evolution of particles ejected under different ejection mechanism, cometary or gravitational escape. Left panel: standard deviation of mean anomalies of simulated particles, $\sigma(M)$; the encircling limit at $\sigma(M)=60^\circ$ follows the definition in \citet{Ye2016a} (described in main text). Right panel: along-orbit distance of the center of the particles (represented by the mean of the mean anomaly, $\overline{M}$) to Phaethon.}
\label{fig:ejecta-evolution}
\end{figure*}

We find that it only takes about 250~years ($\sim200$~orbits) for particles to spread to the entire orbit of Phaethon (Figure~\ref{fig:ejecta-evolution}, left panel). The timescale is much shorter ($\sim30$~years) for $\beta=0.001$ particles under cometary ejection. The definition adopted here is the ``encircling limit'' described in \citet[][\S~3]{Ye2016a}: meteoroids have spread to most of the orbit of the parent when the standard deviation of their mean anomalies $\sigma(M)$ have reached $60^\circ$. The mathematical consideration is that 99.7 per cent of the meteoroids have spread to half-orbit, or $180^\circ$ in mean anomaly, assuming a Gaussian distribution along the orbit. [One may wonder why we do not choose $360^\circ$ in mean anomaly instead of $180^\circ$. The reason is that $\sigma(M)$ becomes highly sensitive to regional over-densities when meteoroids are quasi-evenly distributed along the orbit or $\sigma(M)\rightarrow120^\circ$, as reflected in the figure, and is no longer a good indicator of the dispersion of meteoroids along the orbit.]

Regardless of the driving mechanism and size ($\beta$) of the particle, it only takes a small number of orbits for the bulk of the ejecta to spread to significant distances from Phaethon, ($\sim$ 0.1~au). This means that if any materials larger than typical meteor sizes (see below for more on meteor sized particles) were recently ejected, they could be closer to the Earth than Phaethon itself during their late 2017 approach to the Earth, and are potentially detectable using existing facilities. Examination of the existence of such ejecta will therefore provide a useful diagnosis as to recent activity of Phaethon that is otherwise irretrievable. This will also provide useful information for the design of any Phaethon-bound spacecraft like DESTINY$^+$.

We find that $\beta=0.001$ particles quickly drift away from Phaethon (Figure~\ref{fig:ejecta-evolution}, right panel). This radiation-driven process is very fast due to the extreme orbit of Phaethon, with the along-orbit recession rate being approximately 0.1~au/yr. This means that it takes less than an orbit for small particles to drift to different locations of the orbit and effectively blend into the Geminid background.

\section{Hubble Space Telescope Search} \label{sec:obs}

What type of facility is needed to detect mini-fragments accompanying Phaethon? The Minimum Orbit Intersection Distance (MOID) between the orbits of Phaethon and the Earth is 0.02~au, or 3.5 times smaller than Phaethon's closest distance during its 2017 approach. A fragment several meters across will be in the 25--27 magnitude range at a geocentric distance $\varDelta=0.02$~au, assuming Phaethon-like albedo (0.11), a typical phase coefficient (0.035 mag/deg), and a phase angle at $\sim100^\circ$. We note that phase effect at such large angle is best corrected using more sophisticated model \citep[c.f.][]{Li2015}, but for our purpose as a zeroth order approximation, the linear model is good to $<1$~magnitude at a phase angle of $\sim100^\circ$ \citep[][Fig. 1]{Masoumzadeh2015}, and is very much in line with previous measurement of Phaethon up to $\sim80^\circ$ \citep{Ansdell2014}. We also note that the on-sky position of the MOID point is towards the direction of the Sun, resulting a significant brightness reduction due to phase effect as well as a limited observation window for ground-based observers. The short observation window also makes it difficult to perform long exposures, further limiting the effectiveness of a ground-based search.

We secured two orbits with the Hubble Space Telescope (HST) in order to search for fragments accompanying Phaethon. To guide the search, we use the previously-described numeric model to simulate the on-sky particle density. We stick to the gravitational escape model because existing observations do not show Phaethon as a comet over its recent orbits. We first integrate Phaethon back to the year of 1900, an arbitrarily-chosen time that is far back enough to include all ``recent'' activity, and then integrate it forward, releasing $\beta=0$ particles at each perihelion representing large meteoroids. The orbital elements of Phaethon are extracted from JPL orbital solution \#627, as tabulated in Table~\ref{tbl:orb}. Particles with a distance from Earth $\varDelta \leq 0.02$~au during their 2017 approach are saved for further analysis.

\begin{table}
\begin{center}
\caption{Orbital elements of Phaethon (from JPL orbital solution \#627) and the ``master'' particle used to guide HST observation. Uncertainties of the orbit of Phaethon are very small (at the order of $\Delta q \sim 10^{-8}$~au and $10^{-6}$~deg for angular elements) and are not propagated into the simulation. Reference frame is in J2000 coordinates.\label{tbl:orb}}
\begin{tabular}{lcccccc}
\hline
 & Phaethon & Master particle \\
\hline
Epoch & TDB 2017 Dec. 14.5 & TDB 2017 Dec. 14.5 \\
Perihelion distance $q$ (au) & 0.139897 & 0.139913 \\
Eccentricity $e$ & 0.889954 & 0.889941 \\
Inclination $i$ & $22.256382^\circ$ & $22.272522^\circ$ \\
Longitude of the ascending node $\Omega$ & $265.228487^\circ$ & $265.225814^\circ$ \\
Argument of perihelion $\omega$ & $322.176023^\circ$ & $322.173061^\circ$ \\
Mean anomaly $M$ & $331.226910^\circ$ & $334.156887^\circ$ \\
\hline
\end{tabular}
\end{center}
\end{table}

\begin{figure}
\includegraphics[width=0.5\textwidth]{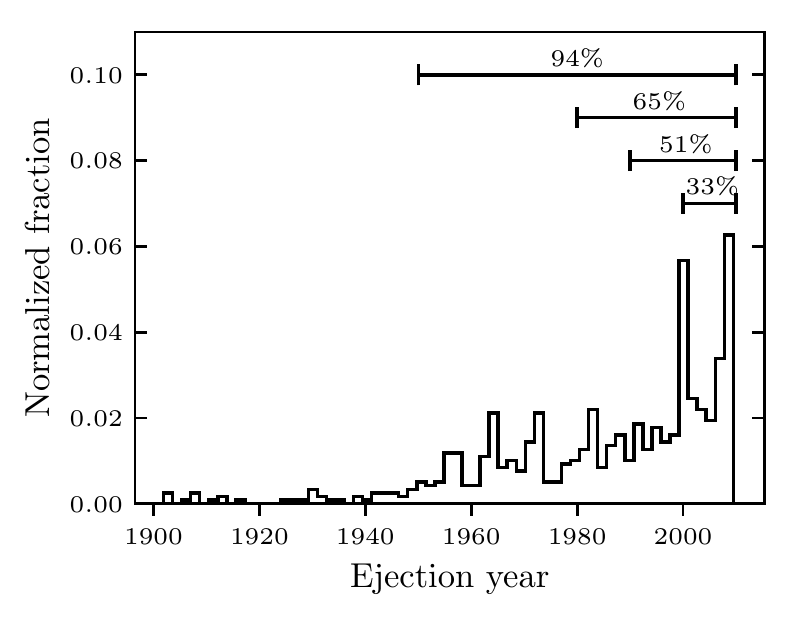}
\caption{Age distribution of the particles that will approach the Earth within 0.02~au during their flyby on 2017 December 14. Contributions of the particles ejected after 1950, 1980, 1990 and 2000 are also given.}
\label{fig:ejecta-age}
\end{figure}

\begin{figure}
\includegraphics[width=0.5\textwidth]{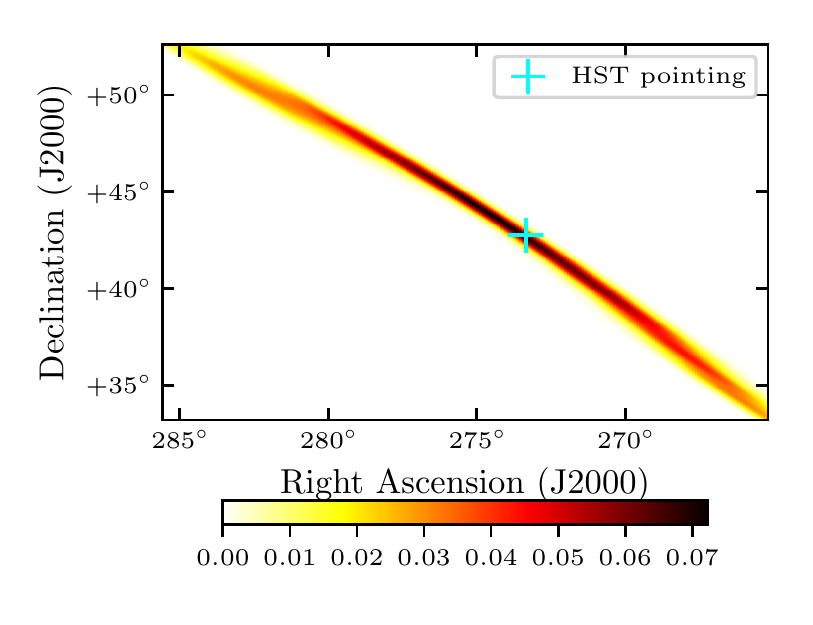}
\caption{On-sky density of the particles that were released after the 1980s and will approach the Earth within 0.02~au during their flyby on 2017 December 14, assuming a geocentric observer with time at UT 18:00, 2017 December 14. Planned HST pointing is marked with a ``+'' sign.}
\label{fig:onsky}
\end{figure}

Figure~\ref{fig:ejecta-age} shows that about most of the $\varDelta \leq 0.02$~au particles in the 2017 approach were released after the 1980s. These particles spread along a $\sim30^\circ$-long arc during their closest approach (Figure~\ref{fig:onsky}). Recognizing that it is impractical to cover the entire arc to 25--27 magnitude in the narrow time window of the closest approach, we decide to use HST to sample the center of this particle swarm, which is equivalently the place with highest particle density as marked on Figure~\ref{fig:onsky}. We refer to the particle that corresponds to this place as the ``master'' particle and tabulate its orbital elements in Table~\ref{tbl:orb}. Apart from the HST, we also secured time on the Gemini North telescopes, the Canada-France-Hawai\textquoteleft i telescope, and other smaller telescopes for a shallower-wider search, which will be described in a forthcoming paper.

We employ the Wide Field Channel (WFC) CCD on the HST Advanced Camera for Surveys (ACS) for our search. ACS/WFC was chosen over WFC3/UVIS because it has 55\% larger field of view, and is more sensitive at redder wavelengths which is important for small body observations. The time of the MOID crossing in 2017 falls at around UT 14:00, 2017 December 14, however the observations were scheduled after 16:30 due to the interference from the South Atlantic Anomaly (SAA).

\begin{figure*}
\includegraphics[width=\textwidth]{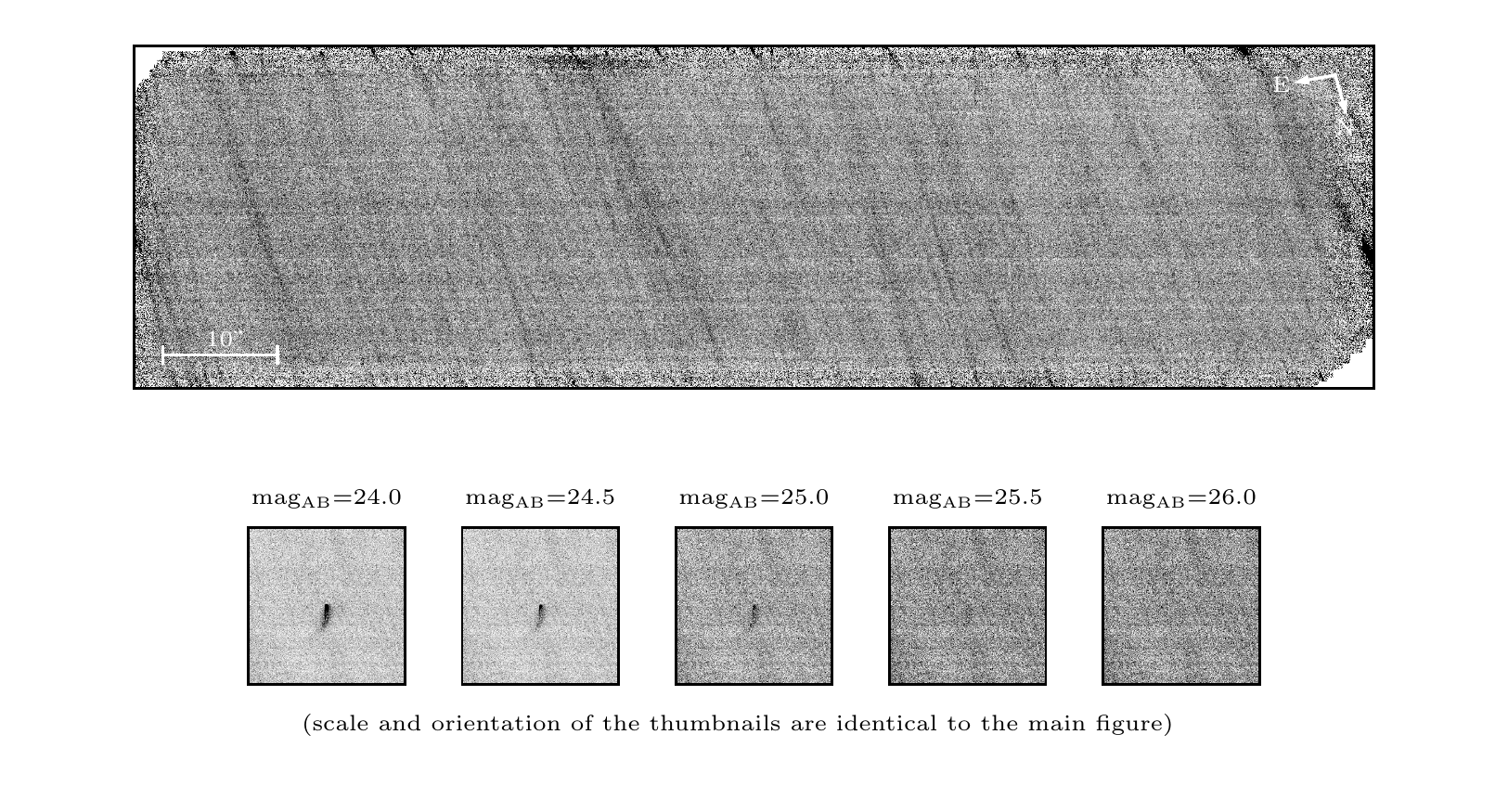}
\caption{Upper panel: combined HST image, with direction of the compass good to $\sim10^\circ$; lower panel: behavior of the synthetic master particle in the combined image.}
\label{fig:hst}
\end{figure*}

The observation was complicated by the strong curvature in the motion of the target as well as the fact that HST was not designed to track in curvature. The work-around was to split the exposure into many shorter exposures so that the target would not smear out too much in each frame. However, since ACS can only hold two full-frame WFC images at a time, we had to use the subarray mode to ensure reaching sufficient depth while sacrificing coverage. As a result, only 1/16 of the WFC/CCD was used for on-sky exposure, while the other 15/16 was used for buffer dumping. A total of 30 exposures was planned over the two orbits, with each 1~min in length. Under this strategy, smearing is reduced to a few arcsec per frame, variable from frame to frame. The F606W filter was used for our observations because it offers maximum sensitivity. Individual frames are median combined, following the motion of the master particle using SWarp \citep{Bertin2010}, as shown in Figure~\ref{fig:hst}.

Since the motion of the imaginary particle is strongly curved, the behavior of such a particle in the stacked image would be sophisticated and needs to be understood. The behavior of the master particle is modeled using the computed ephemerides of the particle and the Tiny Tim software \citep{Krist2011}. This is also useful for determining the depth of the image. We modeled particles of varying brightness at a step of 0.5~mag down to $\mathrm{mag_{AB}}=27.5$ using a solar-type spectrum. We assume the particle falls at the center of the chip and ignore the PSF variation across the chip, since the contribution of such variation to the final PSF of the particle is negligible at the presence of trailing effect due to the motion of the particle. Implanted frames are median combined as what we did for the processing of the original images. In Figure~\ref{fig:hst} we show simulations with brightness from $\mathrm{mag_{AB}}=24.0$ to 26.0. Particles behave like a tailed comet due to the change of the magnitude and the orientation of the smearing effect. From these thumbnails, we estimate that our search is completed to $\mathrm{mag_{AB}} \approx 25$, which translates to a $\sim4$-m diameter object at $\varDelta=0.02$~au, assuming Phaethon-like albedo and accounting for phase effects. We find no obvious candidates for meteoroids from a visual inspection of the stacked image shown in Figure~\ref{fig:hst}. 

\section{Discussion} \label{sec:disc}

What does the HST search imply on the recent activity of Phaethon? The on-sky footprint of our search is 1/16 of the field-of-view of ACS/WFC, that is, about $0.0002~\mathrm{deg^{2}}$. Since no candidate was found, the on-sky particle density is below $1/0.0002=5000~\mathrm{deg^{-2}}$, or $\sim5000 \pm 200~\mathrm{deg^{-2}}$ at $3\sigma$ level, assuming their on-sky distribution follows Poisson statistics. The delivery rate of post-1980 particles into the ``bulls-eye'' zone can therefore be derived from our earlier simulation. 

A critical parameter that defines the ``bulls-eye'' zone is the range of $\varDelta$: real-life fragments at different $\varDelta$ have slightly different motion rates compared to the master synthetic particle, therefore can slowly drift away from the tracking position. The speed that the particles drifts away from each other is essentially the difference in the angular motion rate, $\dot{\omega}$. If we take the speed of the particles perpendicular to our line-of-sight to be $v_{\|}=30~\mathrm{km/s}$ (i.e. the geocentric speed of the Geminid meteoroids), $\dot{\omega}$ can then be approximated by $\dot{\omega}=v_{\|}/\varDelta$. The differential of $\dot{\omega}$ with respect to $\varDelta$ is therefore

\begin{displaymath}
\frac{\mathrm{d} \dot{\omega}}{\mathrm{d} \varDelta} = -\frac{v_{\|}}{\varDelta^2}
\end{displaymath}

\noindent where $\varDelta=0.02$~au is the geocentric distance to the particles in the bulls-eye. Plugging in the numbers, we find $\mathrm{d} \dot{r} \rightarrow 1''$ over a HST orbit when $\mathrm{d} \varDelta \rightarrow 10^{-5}~\mathrm{au}$, therefore the effective range of our search is about $10^{-5}~\mathrm{au}$.

From our simulation, we derive that approximately $0.5\%$ particles ejected since the 1980s ended up in this bulls-eye. In Figure~\ref{fig:onsky} we see that the probabilistic density in the sky where HST pointed to is about 0.07 per square degree. It can be calculated that $0.5\% \times 0.07 = 0.035\%$ of particles will end up in the sky being searched. The $3\sigma$ upper limit of fragments in the Geminid stream placed by our search is therefore $5200/0.035\% \approx 1\times10^{7}$ accumulated over the past $\sim30$ orbits. 

However, our search only provided a single measurement along the mass/size range. The total mass depends on the actual shape of the mass/size distribution which is usually taken to be a power-law distribution. Meteor and lunar impact flash observations suggested that young Geminid meteoroids are dominated by larger meteoroids \citep{Rendtel1999,Blaauw2011,Blaauw2017}, with a differential size distribution index of $\alpha \approx 3.1$ to $3.4$, compared to $\alpha \approx 5$ for the background component (for a size distribution of the form $r^{-\alpha}$). However, meteor and lunar flash observation is only applicable to particles of sizes up to $\sim10$~cm; it is unknown if such power-law extend into meter-class regime. In fact, in-situ observations of near-Earth asteroids suggest a generally steeper slope for 10-m-class boulders, from $\alpha=4.3$ \citep[Itokawa; c.f.][]{Michikami2008,Mazrouei2014} to $\alpha=5.4$ \citep[Toutatis; c.f.][]{Huang2013,Jiang2015}. [Meanwhile, size distributions of fragments from cometary disruptions are in the $\alpha \approx 3 - 4$ range \citep[e.g.][]{Ishiguro2009,Jewitt2016}, but as aforementioned, the formation mechanism for PGC is likely different from the mechanism of typical cometary disruptions.] Since $\alpha=4$ is the ``flipping point'' where mass in distributions with $\alpha<4$ is dominated by largest particles and vice versa, mass of ejecta from Phaethon would be dominated by meter-class ejecta if similar $\alpha>4$ distribution applies to 10-m-sized boulders on Phaethon. Total mass of the ejecta is therefore dependent to where the distribution transitions from $\alpha<4$ to $\alpha>4$. Considering the bulk density of Geminids to be $2900~\mathrm{kg~m^{-3}}$ \citep{Babadzhanov2009}, the upper limit of total mass loss placed by our search is $\sim10^{12}$~kg over the past $\sim30$~orbits if we assume such a transition point occurs at a few meters. This is a small fraction of the estimated total mass of the Geminid meteoroid stream, which is of the order of $10^{13}$ to $10^{16}$~kg \citep[c.f.][and references therein]{Ryabova2017}.

The derived mass loss constraint is $7-8$ magnitudes higher than the periodical dust emission detected by STEREO \citep[$\sim10^4-10^5$~kg, c.f.][]{Hui2017}. However, we caution that the regime of the hypothetical mass loss examined by this work is likely different from the one detected by STEREO: dust emission observed by STEREO is likely caused by thermal fracturing, which is capable of ejecting sub-millimeter-sized dust grains \citep{Jewitt2012}, while the regime that is being investigated by this work should be capable to launch much larger (meter-sized) particles.

In fact, the derived mass loss constraint is more comparable to the estimated mass loss of the mega-outburst of comet 17P/Holmes in 2007 \citep[$10^{10} - 10^{12}$~kg, e.g.][]{Reach2010,Li2011}. Such disruption would have brought the brightness of Phaethon up by a factor of a million, or $\sim15$ magnitudes. Brightness of Phaethon typically varies from 12 (perihelion) to 20 (aphelion). A Holmes-like event will produce prolonged brightening due to the massive amount of dust released, and can take a year to fully subside. Since its discovery on 1983 October 11, Phaethon has been regularly monitored except for the short periods around its perihelion (when Phaethon is only a few degrees from the Sun and cannot be observed from the ground), though the interval between observations before the operation of global Near-Earth Object (NEO) surveillance network (about 1995) can be as long as a few years. Nevertheless, the result of our search is consistent with the fact that no brightness anomaly (apart from the ones recently noticed in STEREO data) has been reported in the past several decades.

The result of our search also suggests that Phaethon-bounded missions are unlikely to encounter dense dust clouds, since no evidence of large recent disruption is found. In addition, Figure~\ref{fig:ejecta-evolution} shows that dust generated by recent disruption quickly move away from Phaethon. As a cursory test, we retrieve the $\beta=0.001$ (millimeter-sized) particles from our simulation and compute the volume density within 0.1~au from Phaethon on 2017 December 14. Assuming $\alpha=3.4$ within the millimeter-meter size regime and a relative speed of $30~\mathrm{km/s}$, the null result of the HST search implies a flux of $<10^{-12}~\mathrm{m^{-2}~s^{-1}}$ for mm-sized dust, comparable to the background meteoroid flux \citep{Grun1985}.

\section{Conclusion} \label{sec:conc}

We present a guided HST search of meter-sized fragments in the vicinity of Phaethon, with the goal of looking for evidence of recent disruption of this object.

We first numerically simulated the motion of a number of virtual particles in order to decide the best pointing for HST. We found that the dynamical evolution of Phaethon-originated particles is quick, possibly due to stronger radiation differentiation resulting from the small perihelion distance of Phaethon. Particles ejected just above escape speed take $\sim250$~yr to encircle the entire orbit, while those ejected via sublimation-driven activities common in typical comets take only $\sim30$~yr.

Our HST search was completed down to 4-meter-class limit assuming Phaethon-like albedo, and was expected to detect 0.035\% particles ejected by Phaethon in the past $\sim4$ decades. The negative result suggests that the total mass loss of Phaethon over the past a few dozen orbits is $<10^{12}$~kg at $3\sigma$ level, a small fraction of the Geminid meteoroid stream, taking the best estimates of size power-law from meteor observations and spacecraft data. Our result also implies a mm-sized dust flux of $<10^{-12}~\mathrm{m^{-2}~s^{-1}}$ within 0.1~au of Phaethon, a level that is comparable to background meteoroid flux. This suggests that any Phaethon-bound spacecraft, such as the DESTINY$^+$ mission, is unlikely to encounter dense dust clouds.

Our work is admittedly limited by the small searched volume restricted by the short observing window. Future facilities with wider field-of-view, such as the Large Synoptic Survey Telescope (LSST) and James Webb Space Telescope (JWST), will enable wider searches over a short period of time and could potentially provide more stringent constraints.




\acknowledgments

We thank an anonymous referee for his/her comments. This work is based on observations made with the NASA/ESA Hubble Space Telescope, obtained at the Space Telescope Science Institute (STScI), which is operated by the Association of Universities for Research in Astronomy, Inc., under NASA contract NAS 5-26555. We thank the STScI staffs, especially Ralph Bohlin and Alison Vick, for making this challenging observation happening. These observations are associated with program 15357. Support for program 15357 was provided by NASA through a grant from the Space Telescope Science Institute, which is operated by the Association of Universities for Research in Astronomy, Inc., under NASA contract NAS 5-26555, to Thomas Prince. P. W. is supported by the Natural Sciences and Engineering Research Council of Canada. M.-T. is supported by a NASA grant to David Jewitt.

\facilities{HST(ACS)}
\software{Matplotlib \citep{Hunter2007}, MERCURY6 \citep{Chambers2012}, SWarp \citep{Bertin2010}, Tiny Tim \citep{Krist2011}}

\end{CJK*}
\bibliographystyle{aasjournal}
\bibliography{ms}



\end{document}